\documentclass{article}

\usepackage{PRIMEarxiv}

\usepackage[utf8]{inputenc} 
\usepackage[T1]{fontenc}    
\usepackage{hyperref}       
\usepackage{url}            
\usepackage{booktabs}       
\usepackage{amsfonts}       
\usepackage{nicefrac}       
\usepackage{microtype}      
\usepackage{lipsum}
\usepackage{fancyhdr}       
\usepackage{graphicx}       
\graphicspath{{media/}}     

\usepackage{subcaption}
\usepackage{multirow} 
\usepackage{amsmath}
\usepackage{diagbox}
\usepackage{siunitx}
\usepackage{comment}
\usepackage{nomencl}

\makenomenclature
\newcommand{\round}[1]{\num[round-mode=places,round-precision=2]{#1}}
\newcommand{\roundto}[2]{\num[round-mode=figures,round-precision=#1]{#2}}
\DeclareMathOperator*{\argmin}{arg\,min}
\usepackage{nomencl}
\title{Reduced order modeling of the unsteady pressure on turbine rotor blades using deep learning

}

\author{
  Dominique J , Salesses L , Thomas J. F , Baert L , Benamara T \\
  Cenaero, Gosselies, Belgium \\
  \texttt{joachim.dominique@cenaero.be} \\
  \and
 \textbf{ Mastrippolito F , Flament T} \\
  Safran Helicopter Engines, Bordes, France \\
}

\begin{document}
\maketitle

\begin{abstract}
In transonic turbine stages, complex interactions between trailing edge shocks from nozzle guide vanes and rotor blades generate unsteady wall pressure fields, affecting the rotor aerodynamic performance and structural integrity. While shock-related phenomena are prominent, unsteady pressure fluctuations can also arise in subsonic regimes, where wake interactions alone are sufficient to induce instationarities. Traditional methods like Unsteady Reynolds-Averaged Navier–Stokes (URANS) simulations, while sufficiently accurate, are computationally expensive. To address this, a novel deep learning based Reduced Order Model (ROM), built upon a database of URANS simualtions, is proposed to predict unsteady pressure fields on a turbine rotor blade at a fraction of the simulation cost. Specifically, the model consists of a Variational Auto-Encoder (VAE) integrated with a Gated Recurrent Unit (GRU) to capture time-series data, addressing the limitations of traditional linear ROMs in capturing efficiently nonlinear phenomena, such as moving shocks. The objective of this work is to develop a ROM capable of accurately reproducing the unsteady pressure fields obtained from URANS simulations while significantly reducing computational costs. The proposed ROM is applied to the Turbine Aero-Thermal External Flows (TATEF2) project configuration, a well-established test case in turbomachinery research that is representative of modern high-pressure turbine stages, particularly in terms of shock-wave interactions and wake dynamics. The model performance is evaluated using a combination of machine learning quality metrics and design-oriented criteria, such as the accuracy of the first harmonic in the Fourier transform of the unsteady pressure field. Additionally, the influence of the simulation database size on model accuracy is analyzed, recognizing that the number of training simulations required to achieve task-specific accuracy is a key constraint on the industrial applicability of such approaches. This preprint article has been accepted for publication in Journal of Turbomachinery. After publication, it will be available at \href{https://doi.org/10.1115/GT2025-151476}{https://doi.org/10.1115/GT2025-151476}
\end{abstract}

\nomenclature{$d$}{Inter row distance [mm]}
\nomenclature{$t$}{Time [s]}
\nomenclature{$T$}{Period [s]}
\nomenclature{$K_k$}{Reduced stiffness matrix}
\nomenclature{$M_k$}{Reduced mass matrix}
\nomenclature{$C_k$}{Reduced damping matrix}
\nomenclature{$\mathbf{x}$}{Pressure field [kPa]}
\nomenclature{$\hat{\mathbf{x}}$}{Reconstructed pressure field [kPa]}
\nomenclature{$z$}{Latent space variables}
\nomenclature{$N_s$}{Number of stator blades}
\nomenclature{$N_r$}{Number of rotor blades}
\nomenclature{$N_u$}{Number of cells along the rotor span}
\nomenclature{$N_v$}{Number of cells along the rotor chord}
\nomenclature{$N_t$}{Number of timesteps}
\nomenclature{$S$}{Number of samples}
\nomenclature{$D_{KL}$}{Kullback-Leibler divergence}
\nomenclature{$r$}{Generalized coordinate}
\nomenclature{$\alpha$}{Flow angle of attack [$^{\circ}$]}
\nomenclature{$\beta$}{Blade pitch angle [$^{\circ}$]}
\nomenclature{$\Pi$}{Stage pressure ratio}
\nomenclature{$\Omega$}{Angular velocity [rad s$^{-1}$]}
\nomenclature{$\zeta$}{VAE loss weighting factor}
\nomenclature{$\theta$}{ROM parameters}
\nomenclature{$\omega$}{Frequency [rad s$^{-1}$]}
\nomenclature{$\gamma$}{Tolerance curve parameter $n^{\circ} 1$}
\nomenclature{$\tau$}{Tolerance curve parameter $n^{\circ} 2$}
\nomenclature{$tol$}{Tolerance curve parameter $n^{\circ} 3$}
\nomenclature{$\phi_k$}{Harmonic deformation mode}
\nomenclature{ANN}{Artificial Neural Network}
\nomenclature{CNN}{Convolutional Neural Network}
\nomenclature{DoE}{Design of Experiments}
\nomenclature{FFT}{Fast Fourier Transform}
\nomenclature{MSE}{Mean Square Error}
\nomenclature{MAE}{Mean Absolute Error}
\nomenclature{MRE}{Mean Relative Error}
\nomenclature{MLP}{MultiLayer Perceptron}
\nomenclature{ROM}{Reduced Order Model}
\nomenclature{GRU}{Gated Recurrent Unit}
\nomenclature{TPM}{Temporal Prediction Model}
\nomenclature{AE}{Auto-Encoder}
\nomenclature{VAE}{Variational Auto-Encoder}
\nomenclature{BPF}{Blade Passing Frequency}
\nomenclature{(U)RANS}{(Unsteady) Reynolds-averaged Navier--Stokes}

\printnomenclature

\section{Introduction}

In highly loaded turbine stages, the exit flow from the nozzle guide vane often reaches transonic speeds, creating a complex interaction between the stator trailing edge shocks and the rotor blades. This interaction significantly impacts the aerodynamic and mechanical performance of the rotor. Understanding and accurately modeling the rotor unsteady wall pressure field is crucial and requires a detailed understanding of the moving shock patterns and fluctuations produced beneath the turbulent boundary layer. Although shock-related phenomena are prominent, unsteady pressure fluctuations can also occur in subsonic regimes, where wake interactions alone are enough to induce instabilities.

The unsteady pressure loading on turbine blades is crucial for assessing both their structural integrity and aerodynamic performance. This loading can be assessed through various CFD modeling approaches, each offering a different level of fidelity. Unsteady RANS methods are commonly preferred over lower-order approaches, such as those based on Euler solvers, due to their ability to capture nonlinear effects of compressibility and viscosity. Additionally, they are often favored in industry over higher-order methods like LES, which, while more accurate, are considerably more computationally demanding. Nonetheless, CFD simulations remain resource-intensive and time-consuming, presenting significant challenges for application across diverse operating conditions under tight time constraints.

To address these challenges, there is a growing interest in leveraging machine learning techniques and Reduced Order Modeling (ROM) to create faster unsteady pressure estimators. These methods can potentially streamline the design process without compromising accuracy. Data-driven ROM, in particular, has gained traction for its ability to reduce the computational cost of CFD models. Combining linear reduction techniques like Proper Orthogonal Decomposition with regression methods such as Gaussian Processes or Radial Basis Functions has shown promise in various applications \cite{winter2016} or optimisation framework \cite{karbasian2022}. However, these approaches often struggle with highly nonlinear phenomena, such as moving shocks pattern, requiring a large number of modes to capture most of the energy spectrum, thereby diminishing their efficiency.

To overcome these limitations, nonlinear ROM approaches utilizing deep learning have been proposed. For example, Hasegawa \textit{et al}. \cite{hasegawa2020} developed a method integrating Auto-Encoders (AE) with Convolutional Neural Network (CNN) layers and recurrent long-short term memory (LSTM) networks to predict unsteady flow fields. This approach demonstrated the practical use of nonlinear ROM for fluid dynamics problems. Similarly, Rosov and Breitsamter \cite{rozov2021} employed a U-net-like architecture to predict static pressure fields around airplane wings, while Jiaobin \textit{et al} \cite{ma2022} used deep neural networks to learn FFT coefficients of unsteady pressure. Hines and Bekemeyer \cite{hines2023} advanced this field further by applying Graph Neural Networks to predict distributed quantities over unstructured meshes. Recently, Solera-Rico \textit{et al} \cite{solera2024} have proposed an approach combining beta-Variational Auto-Encoders (VAE) and transformers on 2D viscous flows. Similarly, advancements in the use of artificial neural networks (ANN) have led to substantial reductions in computational time for turbine blade aerodynamic design optimization \cite{zhang2022, nigro2025journal}.

Our work aims to predict the unsteady pressure field over a turbine rotor blade using VAE trained on a database of URANS simulations. The ability of the VAE to regularize the latent space provides an improvement over traditional Auto-Encoders (AE) \cite{kingma2013}. This VAE is then combined with a Temporal Model (TPM) based on a GRU to predict time-series data. 

To the best of the authors knowledge, this is the first study to apply such an approach to an industrially relevant turbomachinery design, which significantly increases the complexity of the problem compared to previous applications. In addition to demonstrating the feasibility of the nonlinear ROM for complex turbomachinery flows, this work provides an analysis of the impact of training database size on model performance acknowledging that the number of simulations required to achieve task-specific accuracy represents a key constraint for the industrial applicability of such approaches. The model effectiveness will be assessed by its ability to replicate the unsteady static pressure field and the first harmonics of their Fourier transform, which are representative of the forcing terms in the evaluation of Generalized Aerodynamic Forces (GAF) that are used to computed aeroelastic forced response. 

\section{Unsteady flow simulation}

\begin{figure}[htbp]
    \centering
    \includegraphics[width=0.75\textwidth]{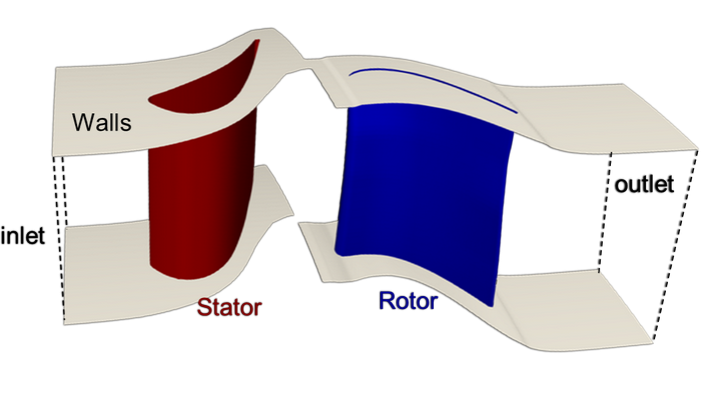} 
    \caption{geometry of the stator and rotor stage}
    \label{fig:geometry}
\end{figure}

The turbine stage geometry analyzed in this study, as depicted in Fig. \ref{fig:geometry}, corresponds to the configuration employed in the Turbine Aero-Thermal External Flows (TATEF2) project, funded by the European Union. The stage consists of a stator with $N_s$ blades and a rotor with $N_r$ blades rotating at an angular velocity $\Omega$ and a blade passing period of $T_{BPF} = \frac{N_s}{\Omega}$. This setup is specifically designed to explore unsteady flow phenomena and interactions under transonic flow conditions. Building on prior research \cite{denos2005, paniagua2008, laumert2002_part1, laumert2002_part2}, which demonstrated the sensitivity of the turbine flow characteristics to various operational parameters such as pressure ratios, inflow angles, and blade spacing, this paper maintains constant blades geometry while examining the blade unsteady pressure field across different wind tunnel operating conditions.

\subsection{Mesh and boundary conditions}

The computational domain is discretized using a structured O-grid mesh around the blades and H-blocks for the rest, comprising $3\times10^6$ cells with refinements focused within the airfoil boundary layer and in the rotor tip clearance. The mesh quality is ensured through standard industrial criteria, including cell skewness, aspect ratio, expansion ratio, and wall distance. Particular attention is given to the mesh on the rotor skin, as this is where the unsteady pressure field will be extracted. The grid on the rotor skin consists of $N_u = 100$ cells along the span and $N_v = 192$ cells along the chord. The mesh is structured but non uniformly distributed with refinement area around the tip, the hub, the leading and trailing edges.

At the inlet, the flow angle is imposed with the total pressure and temperature. At the outlet, static pressure is imposed to match a specific pressure ratio between the inlet and outlet planes. Periodic boundary conditions are applied on both lateral sides, and no-slip wall conditions are imposed at the hub while a thin gap is present between the rotor tip and the upper wall.  The rotor moves at the angular velocity $\Omega$, and the interface between the fixed stator and the moving rotor domain is defined as a rotor-stator interface with so-called chorochronic or phase-lag boundary conditions using 16 harmonics of the BPF \cite{gerolymos2002}.

\subsection{Flow simulation model}

The numerical CFD simulation consists of two main steps. Initially, a steady 3D RANS simulation is performed to initialize the solution. Following this, the unsteady solution is obtained starting from the steady state using URANS until a moving average convergence of the mass flow rate is achieved. These simulations are conducted using the elsA software \cite{cambier2011} with the k-l Smith turbulence model. The entire process can be performed in average in 810 CPU-hours parallelized using 128 CPU on one AMD EPYC 7763 node . 

Once the unsteady simulation reaches periodic convergence, the unsteady pressure field on the rotor cells is extracted for a complete blade passing period of the stator, covering a total of $N_t = 192$ time steps. Additionally, several key quantities are computed to monitor the operating conditions of the turbine stage and ensure the simulation convergence.

\begin{figure}[htbp]
    \centering
    \includegraphics[width=0.5\textwidth]{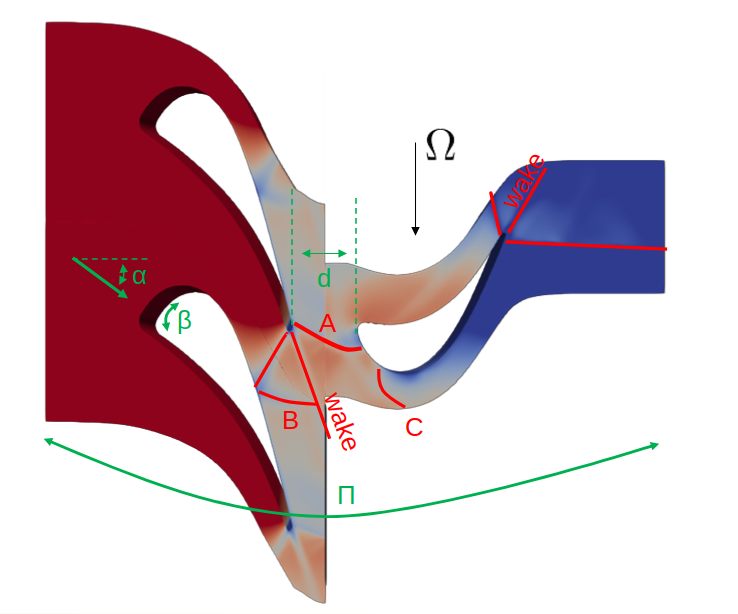}
    \caption{Example of shock patterns in the TATEF2 turbine stage configuration illustrated in red on the static pressure field at mid-span. In green are the design parameters considered for the reduced order modeling database } 
    \label{fig:shockspattern}
\end{figure}

\subsection{Unsteady pressure loading}

To illustrate the complexity of the physics involved, including moving-shocks, we align our analysis with the work of Guillermo Paniagua \textit{et al.} \cite{paniagua2008}. Paniagua \textit{et al.} detailed experimental and numerical analysis of stator-rotor interactions in a transonic turbine stage revealed that multiple shock patterns simultaneously influence rotor loading. When the leading edges of the rotor and stator align, the trailing edge shock from the stator directly impacts the rotor suction side close to the leading edge. A second trailing edge shock reflects off the adjacent stator blade, and the interaction between the rotor turbulent boundary layer and the upstream moving shocks can lead to significant changes in the boundary layer, including the formation of separation bubbles that influence the position of the reflected shocks within the rotor channel.

Similar observations were made in our study, as depicted in Fig \ref{fig:shockspattern}. Several complex flow phenomena involving interactions between turbulent flow, shocks, and boundary layers are apparent. Specifically, oblique shocks originating from the stator trailing edge (Fig \ref{fig:shockspattern} label A) directly impact the rotor leading edge. As the rotor moves, these shocks change positions moving toward the leading edge. Depending on their interaction with the turbulent boundary layer, and the possible induced separation bubble, these shocks create reflections that impact the pressure side of the adjacent rotor blade ( Fig \ref{fig:shockspattern} label C). Another set of oblique shocks (Fig \ref{fig:shockspattern} label B) from the stator trailing edge interact with the rotor only after reflecting off the adjacent stator blade and passing through the wake. Consequently, these shocks are expected to have a lesser influence on rotor loading as compared to shock A. Additionally, depending on the inflow conditions, other shocks may appear within the rotor stage channel.

While such analyses are well established in the context of high pressure transonic turbine stages, it is vital that the ROM can accurately capture the position of the shocks to properly replicate the forces acting on the blades.

\subsection*{Shock tracking using Canny edge detection}

The correct modeling of the shock positions and amplitudes is critical to the evaluation of the stage performance \cite{giles1990} and structural integrity. In order to provide a more quantitative measurement of the shocks positions and to facilitate the analysis of shock patterns, we propose using the Canny edge detection technique \cite{canny1986}. This method, known for its simplicity and efficiency in tracking non-linearities or discontinuities, is particularly useful for visualizing moving shock patterns and was previously employed in several experimental campaign using Schlieren or other shadowgraph methods or even in CFD simulations \cite{fujimoto2019,znamenskaya2021,doroshchenko2023}. Canny edge detection involves several key steps: 1) Gaussian Filtering : Generally, the image is passed through a Gaussian filter to smooth it and reduce noise. However, since we assume the numerical simulation data is noise-free, this step is omitted in our analysis. 2) Gradient Computation: The image is then processed using Sobel filters to compute the gradients, highlighting areas of rapid intensity change. 3) Non-Maximum Suppression: This step identifies the local maxima of the gradients, pinpointing the precise locations of edges. 4) Double Thresholding: Gradients are classified into strong and weak edges using a double threshold set by the user. Strong edges are those with gradients above the high threshold, while weak edges are those between the high and low thresholds. 5) Edge Tracking by Hysteresis: This final step ensures that all strong edges are retained, and only weak edges directly connected to strong edges are preserved. For our specific application, hysteresis is performed in both time and space to effectively track the displacement of shocks over time.

\begin{figure*}[ht!]
    \centering
    \begin{subfigure}{0.33\textwidth}
    \includegraphics[width=\textwidth, trim=50 50 50 50, clip]{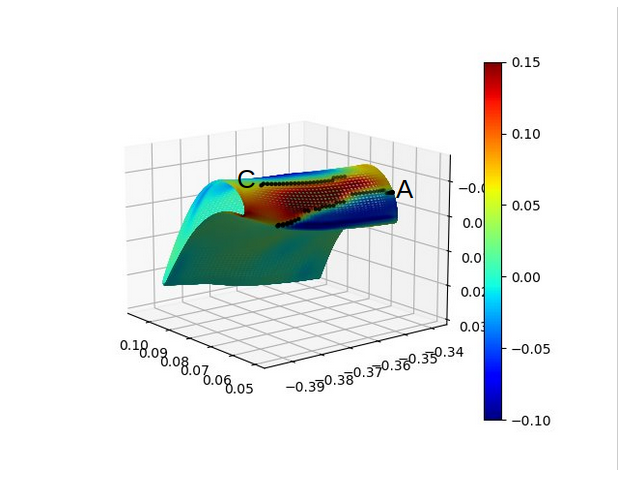}
    \caption{$t = t_s$}
    \label{fig:canny1}
    \end{subfigure}
    \hfill
    \begin{subfigure}{0.33\textwidth}
    \includegraphics[width=\textwidth, trim=50 50 50 50, clip]{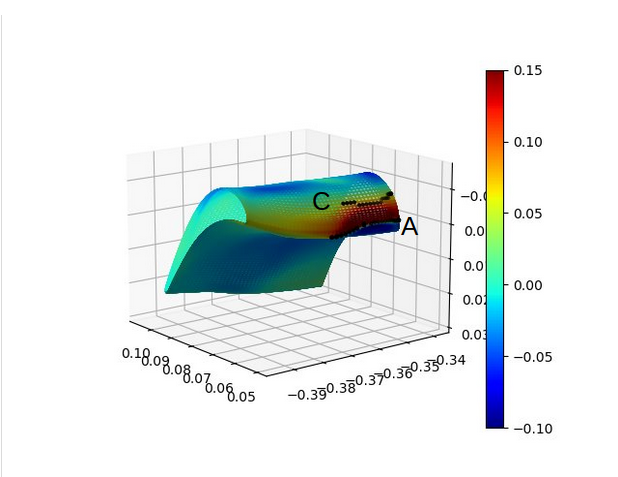}
    \caption{$t = t_s + \frac{1}{5} T_{BPF}$ }
    \label{fig:canny2}
    \end{subfigure}
    \hfill
    \begin{subfigure}{0.33\textwidth}
    \includegraphics[width=\textwidth, trim=50 50 50 50, clip]{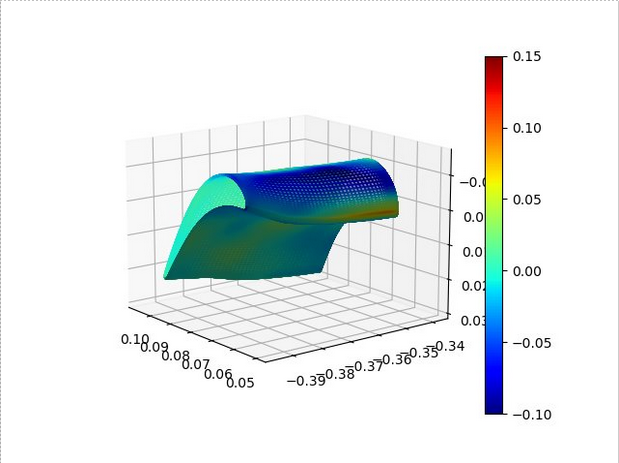}
    \caption{$t = t_s + \frac{2}{5} T_{BPF}$ }
    \label{fig:canny3}
    \end{subfigure}
    \\
    \begin{subfigure}{0.33\textwidth}
    \includegraphics[width=\textwidth, trim=50 50 50 50, clip]{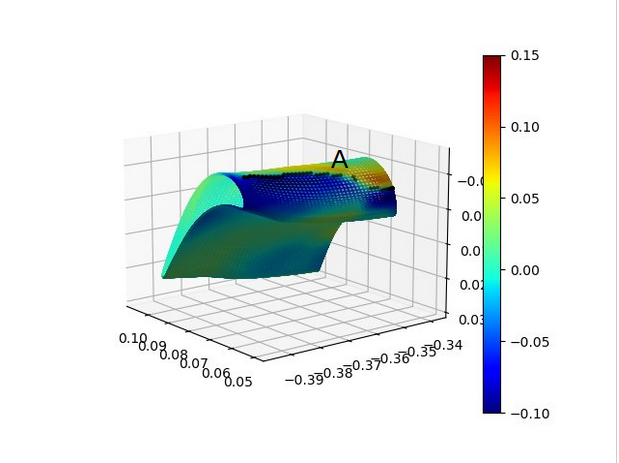}
    \caption{$t = t_s + \frac{3}{5} T_{BPF}$ }
    \label{fig:canny4}
    \end{subfigure}
    \hfill
     \begin{subfigure}{0.33\textwidth}
    \includegraphics[width=\textwidth, trim=50 50 50 50, clip]{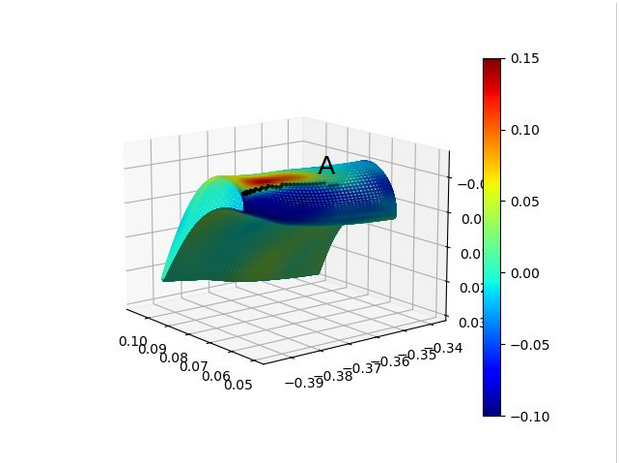}
    \caption{ $t = t_s + \frac{4}{5} T_{BPF}$ }
    \label{fig:canny5}
    \end{subfigure}
    \hfill
    \begin{subfigure}{0.33\textwidth}
    \includegraphics[width=\textwidth, trim=50 50 50 50, clip]{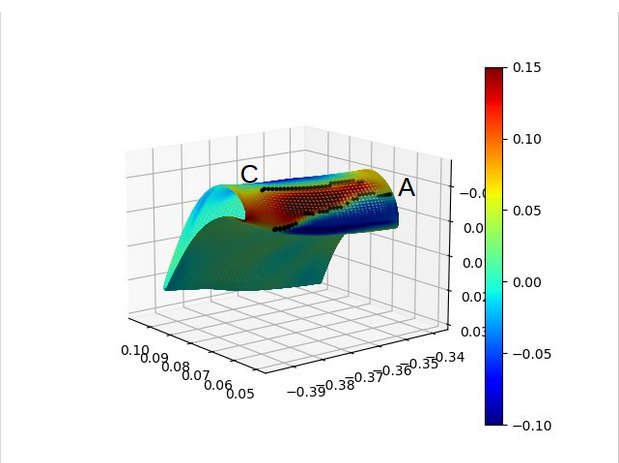}
    \caption{$t = t_s + T_{BPF}$ }
    \label{fig:canny6}
    \end{subfigure}
    \caption{Evolution of the rotor static pressure $p_s $ with the Canny discontinuities in black}
    \label{fig:canny}
\end{figure*}

By employing the modifications introduced in step 1) and 5) to the traditional Canny algorithm, we enhance its suitability for analyzing the temporal and spatial dynamics of shock patterns in turbine stages. This approach allows for precise visualization and tracking of the moving shocks, providing deeper insights into their behavior and interactions. Indeed, we can observe in Fig. \ref{fig:canny1}  the impact on the rotor of shock A and its reflection C. As the rotor continues its rotation, the shock A moves toward the leading edge and the relative position of its reflection changes in Fig \ref{fig:canny2}. This progression continues until the shock passes beyond the rotor blade, as depicted in Fig. \ref{fig:canny3}. Finally, in Fig. \ref{fig:canny4} to \ref{fig:canny6}, we start to observe the impact of shock A from the adjacent stator blade. In addition to this various shocks pattern, it can be shown that the relative position of the shocks on the rotor surface changes during the blade passing period. This illustrates well the complexity of the physics to be considered for the reduced order modeling of the unsteady pressure field. However, it is worth mentioning that the Canny edge detection method may detect non-linearities in the wall surface pressure that are not necessarily caused only by shocks, but appears as a consequence of flow separation or sudden geometrical changes.

\subsection{Generalized Aerodynamic Forces} \label{section:GAF}

Pressure fluctuations on the wall underneath a turbulent boundary layer significantly influence fluid-structure interactions, impacting fatigue-related component failures, transmission of vibro-acoustic noise and its aeroelactic deformation. The importance of fluid-structure coupling is dictated by the spatial and temporal features of the pressure field that develops over the airfoil surface and its interaction with the blade structural response.

This study focuses on the aeroelastic fluid-structure coupling. Assuming a weak coupling between the blade structure and the aerodynamic excitation, and under the hypothesis of linear behavior of the blade structure with harmonic motion, the structural equilibrium can be expressed in a modal formulation. This involves the introduction of generalized matrices and generalized aerodynamic forces (GAF), described by the equation \cite{dowell2021modern}:

\begin{equation}
    r (\mathbf{K}_k - \omega^2_k \mathbf{M}_k + i \omega \mathbf{C}_k) = \mathbf{\phi}_k \mathbf{f}_k
\end{equation}

 where $\mathbf{K}_k$, $\mathbf{M}_k$ and $\mathbf{C}_k$ represent the reduced stiffness, mass, and damping matrices of the blade mechanical system, respectively, projected onto the harmonic deformation mode $\mathbf{\phi}_k$. The parameter $r$ is the generalized coordinate written in the complex formalism that characterizes the amplitude of the deformations and their phase. The right-hand side of the equation corresponds to the GAF. Typically, the GAF is calculated by monitoring the unsteady static pressure on the blade until periodic behavior is established. The forces on the blade are then reconstructed by multiplying the pressure with the normal vector and projecting it onto the deformation mode, which is obtained through mechanical simulations using a finite element solver.

Alternatively, from a purely aerodynamic perspective, the key forcing term is the amplitude of the first harmonic of the Fourier transform applied to the pressure fluctuations. This approach will be preferred in this work. To quantify the uncertainty of this metric, the results were compared against an experimentally validated tolerance curve which permits larger deviations in regions where the amplitude of the first harmonic is particularly low and thus less impactful \cite{SafranHE}: 

\begin{equation}\label{eq:tolerance}
    \frac{|FFT_1^{ROM} - FFT_1^{CFD}|}{FFT_1^{CFD}} \leq tol + e^{-\gamma\left(y-\tau\right)}
\end{equation}

where $\gamma$ , $\tau$ and $tol$ are experimentally validated parameters and $y= |FFT_1^{CFD}|/\max(|FFT_1^{CFD}|)$.  This tolerance curve is shown in Fig. \ref{fig:tolerance} and will be further discussed in Sec. \ref{sec:results}. At the moment, it suffices to say that the amplitude of the harmonic of the fluctuating pressure is an interesting metric to evaluate the performance of the surrogate models presented in next section. Accurately approximating the harmonics of the fluctuating pressure demonstrates the ability of the ROM to reproduce realistic pressure excitations regardless of the deformation modes and the hypothesis used for its computation. This enhances the predictive capability of the surrogate models and demonstrates that they are accurate enough to support advanced post-processing of their predictions.

\begin{figure*}[ht!]
\centering
\begin{subfigure}{0.2\textwidth}
\includegraphics[width=\textwidth, trim=0 0 0 3, clip]{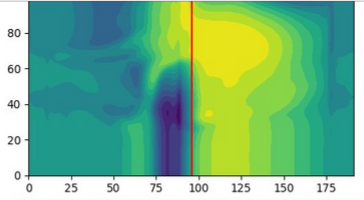}
\caption{Original unfolded image of static pressure field. The red line indicates the leading edge position}
\label{fig:VAE1}
\end{subfigure}
\hfill
\begin{subfigure}{0.5\textwidth}
\includegraphics[width=\textwidth, trim=0 0 0 0, clip]{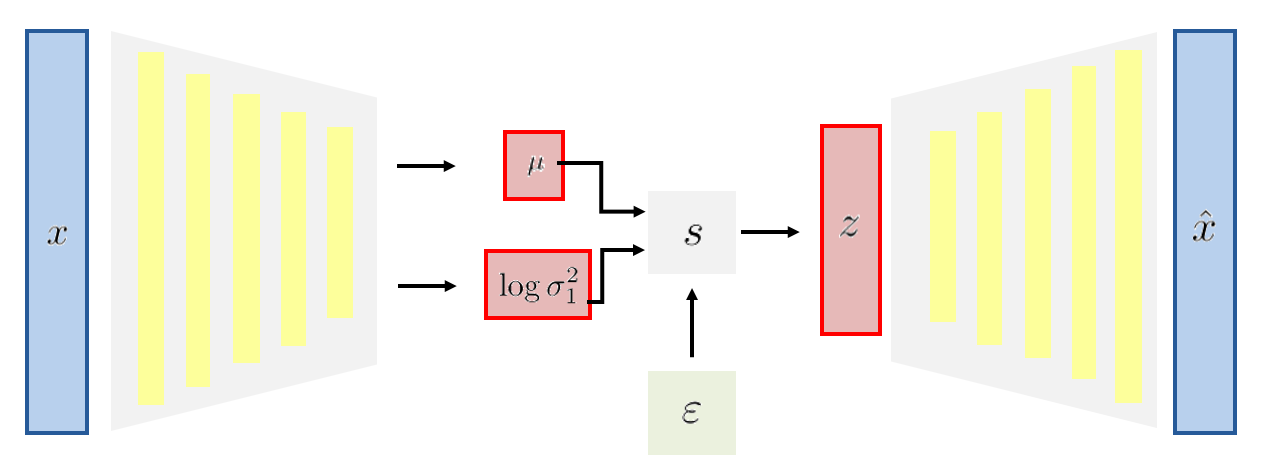}
\caption{Structure of the VAE with $p$ being the input field and $\hat{p}$ being the reconstructed field}
\label{fig:VAE3}
\end{subfigure}
\hfill
\begin{subfigure}{0.2\textwidth}
\includegraphics[width=\textwidth, trim=3 0 0 3, clip]{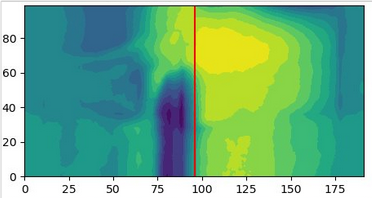}
\caption{Reconstructed unfolded image of static pressure field. The red line indicates the leading edge position }
\label{fig:VAE2}
\end{subfigure}
\\
\vspace{1cm}
\begin{subfigure}{0.8\textwidth}
\includegraphics[width=\textwidth, trim=3 0 0 3, clip]{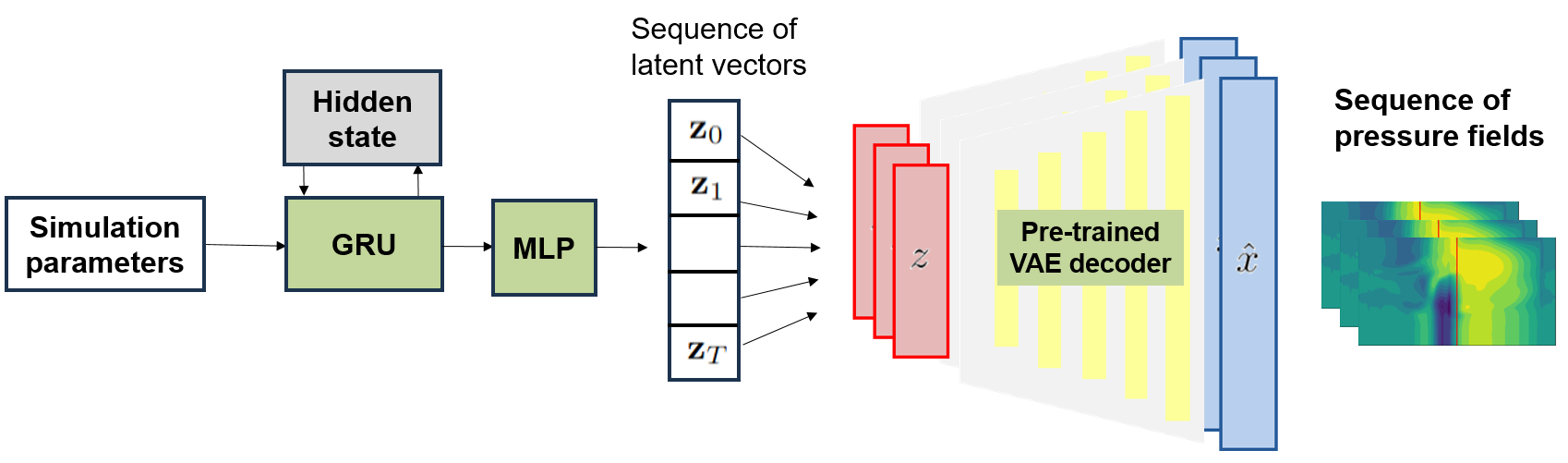}
\caption{Temporal prediction model}
\label{fig:VAE4}
\end{subfigure}

\caption{Global architecture of the surrogate model for unsteady pressure predictions}
\label{fig:VAE}
\end{figure*}

\section{Reduced order Modeling} \label{section:ROM}

We focus on developing a ROM to predict the unsteady, high-dimensional pressure field with moving nonlinearities on a rotor blade. The methodology for constructing this model involves three main steps. First, a database of CFD simulations is generated and partitioned into training, validation, and testing sets. Next, the model architecture is designed and trained. The architecture is structured to decouple spatial and temporal prediction tasks. Spatial predictions are addressed through dimensionality reduction techniques that map the high-dimensional pressure field to a lower-dimensional latent space, while temporal predictions are made by forecasting the trajectory of multiple latent space vectors. Finally, the model performance is evaluated.

\subsection{Numerical dataset} \label{section:numdataset}

The design parameters for the numerical dataset used in this work include the inflow angle, pitch angle, inter-row distances, and stage pressure ratio, denoted respectively as $[\alpha, \beta, d, \Pi]$ and illustrated in Fig. \ref{fig:shockspattern}. The design space spanned by these parameters is sampled using a design of experiments (DoE) approach with Latinized centroidal Voronoi tessellation (LCVT) \cite{romero2006}, facilitated by the in-house Cenaero software MINAMO \cite{sainvitu2010,beaucaire2019}. 

Four distinct datasets were created consisting of 20, 40, 80 and 160 simulations (see Tab. \ref{tab:datasetssize}). They are respectively designated as \textit{small} (S), \textit{medium} (M), \textit{large} (L) and \textit{extra large} (XL). Each dataset is partitioned into two subsets: training and validation. The training subset is used to optimize the model parameters, while the validation set is used to determine stopping criteria to prevent overfitting, utilizing an early stopping method \cite{zhang2005}. Additionally, a separate dataset containing 40 simulations, designated as \textit{Test} (T), was created to serve as a testing set. This dataset, entirely distinct from all the training and validation subsets, ensures an unbiased evaluation of model performance and facilitates fair comparisons across models. 

The LCVT approach ensures efficient exploration of the parameter space by distributing the sampling points in a way that maximizes the coverage of the design space relative to the number of points within the DoEs. Qualitatively, manifold learning methods such as t-SNE can be employed to visualize in 2D the lack of clustering among the training data \cite{van2008visualizing}. Furthermore, the uniformity of the input parameters for the training subset can be verified by examining their probability distribution, ensuring they are uniformly sampled across all DoEs. However, due to the smaller sample size within the validation set and the stochastic nature of their selection, uniform sampling may not always be guaranteed for the validation set.

\begin{table*}[ht]
\centering
\begin{tabular}{|c|c|c|c|c||p{1.5cm}|p{1.5cm}|c|}
\hline
\diagbox{Dataset}{Subset} & Training & Validation & Test & Total &  \multicolumn{2}{c|}{Training time [GPU-h]} & Time per simulation \\
\cline{6-7}
&  &  &  &  & VAE & TPM & [CPU-h]\\
\cline{1-8}
\textit{Small} (S) & 15 & 5 & & 20 & 0.16 & 0.01 & \multirow{4}{*}{810}\\
\cline{1-7}
\textit{Medium} (M) & 32 & 8 & & 40 & 0.29 & 0.04 &\\
\cline{1-7}
\textit{Large} (L) & 68 & 12 & & 80 & 0.46 & 0.11 &\\
\cline{1-7}
\textit{Extra Large} (XL) & 136 & 24 & & 160 & 0.82 & 0.29 & \\
\cline{1-7}
\textit{Test} (T) & & & 40 & 40 & & &\\
\hline
\end{tabular}
\caption{Number of simulations for the different and independent datasets}
\label{tab:datasetssize}
\end{table*}

To enable effective use of pressure data in machine learning models, it is preferable to map the pressure fields defined on the rotor blade surface onto a uniform, rectangular grid of shape $[N_u, N_v]$, resulting in an image representation of the spatial pressure distribution, as shown in Fig. \ref{fig:VAE1}. This mapping is performed without any interpolation as the structured mesh points are simply unfolded onto a plane by removing the tip of the rotor blade and cutting the blade at the trailing edge. As a result, the pixels on opposite sides of the trailing edge are treated as independent and are not considered correlated within the convolutional layers. This approach simplifies the mapping process but assumes that discontinuities at the trailing edge do not introduce significant artifacts in the learned features.

Such image format facilitates efficient processing using standard deep learning methods, such as Convolutional Neural Networks (CNN). Consequently, the entire sequence of pressure fields is represented as a tensor with shape $[N_t, N_u, N_v]$.  However, this approach does not preserve point-to-point distances between the original and mapped representations.
Given the high dimensionality of the pressure field data, employing a reduced-order model is essential to simplify the modeling process and manage computational complexity effectively.

\subsection{Model architecture}

The architecture of the model is designed to decouple spatial and temporal predictions tasks. A VAE model is employed to construct a reduced-order representation of the blade surface pressure field independently of time, mapping it to a lower-dimensional latent space. Temporal prediction, on the other hand, is achieved by forecasting the trajectory of multiple latent space vectors using a TPM built upon GRU. Each latent vector is subsequently decoded into pressure fields through the pre-trained VAE decoder, resulting in a temporal sequence of pressure fields. The latent space regularization within the VAE is crucial and ensures that similar pressure fields have encoded latent representations close in the latent space. This regularization facilitates smooth navigation of the temporal model through the latent space when predicting sequences of encoded pressure field representations. The models were developed in Python within an in-house code built upon the PyTorch library \cite{pytorch}.

\subsubsection{Variational auto-encoders}

VAEs are advanced generative models that leverage unsupervised learning techniques to encode and reconstruct data while simultaneously learning a probabilistic representation in a latent space. Specifically, VAEs are employed as dimensionality reduction tools to project high-dimensional data, such as pressure fields, into a lower-dimensional latent space, where the data can be more efficiently analyzed and generated \cite{kingma2013}.

A defining feature of VAEs is their incorporation of probabilistic modeling within the latent space, which introduces a regularization that traditional autoencoders lack. Thus overcoming known limitation of traditional autoencoders. Specifically, traditional autoencoders may struggle with the issue that data points which are close together in the latent space can be quite different in the high-dimensional space, and interpolated points in the latent space may not correspond to realistic data. The variational approach enforces a specific distribution on the latent variables to ensure meaningful and smooth transitions between different data points.

In VAEs, it is assumed that the latent space follows a Gaussian distribution with a known prior, typically a normal distribution $\mathcal{N}(0,1)$. Consequently, the encoder network, which maps input data to the latent space, outputs two vectors for each pressure field at a time $t_i$: the mean vector $\mathbf{\mu}$ and the logarithm of the variance vector $\log(\mathbf{\sigma}^2)$. These vectors parameterize the Gaussian distribution from which the latent variables are sampled. The regularization of the latent space is performed by minimizing the difference between the posterior probability distribution of the encoder and the prior distribution, ensuring that the learned latent space remains structured and meaningful.

To generate a latent vector $\mathbf{z}$ for use in the decoder, VAEs utilize a technique known as the reparameterization trick. Instead of sampling directly from the Gaussian distribution (which is not differentiable and would hinder gradient-based optimization), VAEs sample from a standard normal distribution $ \epsilon \sim \mathcal{N}(0,1) $ and then transform this sample using the parameters provided by the encoder: $z = \mathbf{\mu} + \epsilon \mathbf{\sigma} $. This transformation allows the model to maintain differentiability and ensures that gradients can be propagated through the sampling process during training of the neural network using typical tools for backpropagation.

The reparameterization trick effectively decouples the stochastic sampling process from the network parameters, enabling the use of backpropagation to optimize the network. The decoder network then takes this latent vector $z$ as input to reconstructs the original data. The VAE model parameters are optimized with two objectives. The first objective is to minimize the reconstruction loss between the original and reconstructed data, measured as the mean-squared error (MSE) on the pressure field. The second objective is to align the learned latent distribution closely with the prior Gaussian distribution, achieved by minimizing the Kullback-Leibler divergence ($D_{KL}$) between the learned latent distribution and the prior. This approach yields the following optimization problem for VAE training:
\begin{equation}
    \mathbf{\theta}^* = \argmin_{\theta} \left(\text{MSE}(\mathbf{x}, \hat{\mathbf{x}}) - \zeta . D_{KL} \left[q_\theta(\mathbf{z} | \mathbf{x}) || p(\mathbf{z})\right] \right)
\end{equation}
where $\mathbf{x}$ denotes the original pressure field, $\hat{\mathbf{x}}$ is the reconstructed pressure field, and $\theta$ represents the VAE model parameters (encompassing both encoder and decoder). Here, $p(\mathbf{z})$ is the prior latent distribution, and $q_\theta(\mathbf{z} | \mathbf{x})$ is the parametrized posterior distribution. The term $\zeta$ is a weighting factor that balances the reconstruction accuracy and the latent space regularization imposed by the KL divergence \cite{higgins2017,burgess2018understanding}. The parameter $\zeta$ must be chosen carefully: it should be sufficiently large to regularize the latent space, ensuring accurate predictions of coherent time sequences for the encoded pressure field representations, yet not so large that it significantly degrades the reconstruction accuracy. Based on empirical testing, a value of $\zeta=10$ was found to yield satisfactory results. The mean-squared error (MSE) between ground truth $\mathbf{x}$ and predictions $\hat{\mathbf{x}}$ is given by
\begin{equation}
    \text{MSE}(\mathbf{x}, \hat{\mathbf{x}}) = \frac{1}{S N_u N_v} \sum_{i=1}^{S} \sum_{u=1}^{N_u} \sum_{v=1}^{N_v} (x_{i, u, v} - \hat{x}_{i, u, v})^2 \label{eq:mse}
\end{equation}
where $S$ is the number of samples and $\mathbf{x}$, $\hat{\mathbf{x}}$ are static pressure tensors with shape $(S, N_u, N_v)$.

To capture the spatial correlations present in the static pressure field, the encoder and decoder are constructed using multiple convolutional layers in Fig. \ref{fig:VAE3}. Strided convolutions are employed in the encoder to downsample the input fields, progressively reducing their dimensions to match that of the latent space. Conversely, strided transpose convolutions are used in the decoder to upsample the latent vector back to the original input field size.
The encoder network, which takes as input the spatial pressure field, consists of five downsampling convolutional layers with $3x3$ kernels, each followed by a LeakyReLU activation function. A linear layer then produces two vectors that parameterize the Gaussian distributions in the latent space. 
The latent space dimensionality is set to 128, determined empirically to balance a 100-fold compression of the input field with sufficient capacity to learn disentangled representations of high-dimensional data.
The decoder network applies an initial linear transformation to the latent input vector, followed by five upsampling convolutional layers with $3x3$ kernels, each interleaved with LeakyReLU activations, ultimately outputting the reconstructed spatial pressure field.

\subsubsection{Temporal prediction model}

To predict the unsteadiness of the pressure field, we employed a specialized neural network architecture known as GRU. GRUs are a type of recurrent neural network (RNN) designed to handle sequences of data by maintaining a hidden state that evolves over time, making them well-suited for modeling temporal dependencies in long sequential data \cite{chung2014empirical, dey2017gate}.

The GRU works by utilizing a gating mechanism that controls the flow of information through the network, allowing it to capture long-term dependencies in the input sequence while mitigating issues like vanishing gradients that commonly affect traditional RNN. The GRU architecture consists of two primary gates: the update gate and the reset gate. The update gate determines how much of the previous hidden state should be carried forward to the current time step, while the reset gate decides how much of the past information should be forgotten. This allows the GRU to effectively manage the balance between preserving important information and discarding irrelevant data over time.

Although the GRU was selected as the preferred approach in this work, previous studies have demonstrated the effectiveness of other temporal models, such as LSTM \cite{hasegawa2020} and Transformers \cite{solera2024}, within similar frameworks. A comparative performance analysis could potentially rank these approaches against each other. However, the focus of this paper is to develop a functional approach for predicting unsteady pressure fields, rather than conducting an extensive performance comparison between the different methods for predicting timeseries. Therefore, such a comparison is beyond the scope of the current study but could be addressed in future research.

The hidden state vector in a GRU functions as a memory cell, enabling the model to predict subsequent steps based on accumulated information from previous time steps, which is encoded within the hidden state. For the first step of a temporal sequence, the hidden state is typically initialized to zeros or ones, as there is no prior information about preceding steps. However, this initial state, lacking any prior knowledge, may introduce artifacts in early predictions, especially when the model is trained on a limited amount of data. More advanced initialization techniques can mitigate this issue, for example using noise injection to the hidden state vector \cite{zimmermann2012forecasting}. In this work, an optimized hidden state initialization method is proposed that leverages the periodic nature of the pressure signal. Specifically, data augmentation is used to correctly initialize the hidden state. The GRU is executed over two consecutive, identical periods of the signal by repeating the training data. During the first pass, the hidden state is initialized to zeros, and the GRU processes one full period $T_{BPF}$ to generate a hidden state that encapsulates meaningful temporal information. This updated hidden state is then used to initialize the GRU for the second pass over the same period, and the predictions from this second pass are retained as the final model output. This two-pass approach improves prediction accuracy, particularly for the initial time steps of the second pass. Empirical testing confirmed that a single period is sufficient to properly initialize the latent space and improve the early time-step predictions. Although this method utilizes the periodic behavior of the signal to provide better initialization, it does not impose periodicity on the time series predictions themselves. More generally, the methodology proposed in this study could be applied to non-periodic signals, using another initialization procedure.

The full temporal prediction model, designated as TPM, takes as input the simulation parameters and predicts the temporal sequence of pressure fields. It consists of several components as illustrated in Fig. \ref{fig:VAE4}. First, a single GRU layer takes as input the simulation parameters and outputs a sequence of vectors, each of size 128. Each output vector is then processed through a multilayer perceptron (MLP), producing latent space vectors that represent the high-dimensional pressure field in a reduced-dimensional space, as encoded by the previously trained VAE. This sequence of latent vectors is then passed through the pre-trained VAE decoder to reconstruct the corresponding time series of static pressure fields. The temporal prediction model is optimized by minimizing the mean-squared error loss for temporal sequences, between the predicted and the true pressure sequence, and defined by:
\begin{equation}
    \text{MSE}(\mathbf{x}, \hat{\mathbf{x}}) = \frac{1}{S N_t N_u N_v} \sum_{i=1}^{S} \sum_{t=1}^{N_t} \sum_{u=1}^{N_u} \sum_{v=1}^{N_v} (x_{i, t, u, v} - \hat{x}_{i, t, u, v})^2
    \label{eq:mse-t}
\end{equation}
where $S$ is the number of samples and $\mathbf{x}$, $\hat{\mathbf{x}}$ are tensors with shape $(S, N_t, N_u, N_v)$.
During the model training, only the parameters of the GRU and MLP are updated, while the parameters of the VAE decoder remain fixed.

A single GRU layer is sufficient, given the short sequence length and the well-defined temporal pattern. The MLP that follows the GRU consists of three layers, each with 128 hidden neurons, followed by a LeakyReLU activation function. This design adds more optimizable parameters and more complexity to the temporal prediction model, allowing to map more efficiently the GRU output to the latent space.

\subsection{Training and testing} \label{section:traintest}

As the architecture of the model is designed to decouple spatial and temporal predictions tasks, the VAE model and TPM are trained separately. The VAE is trained first as its decoder block is reused directly within the TPM to decode the latent vectors to pressure fields. The decoupling of the training phases ensures that each component of the network was optimized for its specific task, optimizing the VAE to account for space correlation and the TPM for time correlation, leading to more accurate and reliable predictions of the unsteady pressure field. This approach enhanced the overall performance of the network by leveraging the strengths of both the VAE and the TPM.

The VAE was trained using the Adam gradient descent algorithm, which combines momentum with adaptive learning rates by utilizing estimates of both the first- and second-order moments of gradients to efficiently update model parameters. The learning rate was set to $10^{-4}$ with a batch size of 32 samples. The temporal prediction model was also trained using the Adam gradient descent algorithm with a learning rate of $10^{-4}$. The batch size was set to 2 for the small dataset, 4 for medium and large datasets, and 6 for the extra large dataset. 

Both models were trained using an early-stopping approach to prevent overfitting and ensure stable convergence. During training, the loss on both the training and validation sets was monitored at each epoch, and training was stopped when the validation loss failed to improve for a predefined number of epochs, set to 15 for the VAE and 25 for the GRU. Additionally, all input features were normalized using a min-max scaling procedure to ensure numerical stability and improve optimization efficiency by mapping values to a standardized range.

The temporal prediction model counts 101K parameters while the VAE model has 1.1M. Both models were trained on a single A100 GPU, with average training times across different datasets reported in Tab. \ref{tab:datasetssize}. The training of the VAE model is approximately 3 to 10 times longer than the training for the temporal prediction model. The primary challenge lies in coherently compressing the pressure fields into a low-dimensional space. Given that pressure fields at consecutive timesteps are similar, and this similarity is preserved in the latent space due to the regularization of the latent space, the task of the temporal prediction model is facilitated and primarily consists to transition between adjacent state within this latent structured space. Thus, the VAE plays a critical role and carries most of the complexity of the model, resulting in a higher parameter count and an extended training time. It should be noted that training times are minimal compared to data set generation. The latter representing the primary cost in developing a reduced-order model of unsteady pressure.

The metrics of both models are evaluated on each dataset to enable comparison of models performance relative to the training dataset size. Significant variability in neural network performance can arise from the random initialization and the stochastic nature of batch sampling. Training multiple models mitigates this variability and enables a more accurate assessment of model performance. For each dataset, the VAE and TPM models were trained 50 times. This allows to compare the performances in terms of mean and 95\% confidence interval for each metric, as reported in Tab. \ref{tab:metrics_test}. In addition, the model showing the best performance in terms of MSE is retained for reporting on each data set. 

The quality of both the VAE and the temporal prediction models is evaluated using the following metrics evaluated on the single \textit{test} dataset, enabling fair comparisons between all models trained with different datasets.

\begin{itemize}
    \item \textit{Mean-Squared Error} (MSE): defined in Eq. \ref{eq:mse-t}.
    \item \textit{Mean Absolute Error} (MAE):
    \begin{equation}
         \text{MAE}(\mathbf{x}, \hat{\mathbf{x}}) =  \frac{1}{S N_t N_u N_v} \sum_{i=1}^{S} \sum_{t=1}^{N_t} \sum_{u=1}^{N_u} \sum_{v=1}^{N_v} |x_{i, t, u, v} - \hat{x}_{i, t, u, v}|
        \label{eq:mae}
    \end{equation}
    \item \textit{Mean Relative Error} (MRE):
    \begin{equation}
        \text{MRE}(\mathbf{x}, \hat{\mathbf{x}}) =  \frac{1}{S N_t N_u N_v} \sum_{i=1}^{S} \sum_{t=1}^{N_t} \sum_{u=1}^{N_u} \sum_{v=1}^{N_v} \frac{|x_{i, t, u, v} - \hat{x}_{i, t, u, v}|}{x_{i, t, u, v}}
        \label{eq:mre}
    \end{equation}
    \item \textit{Mean Max Error}:
    \begin{equation}
        \text{Mean Max Error}(\mathbf{x}, \hat{\mathbf{x}}) = \frac{1}{S} \sum_{i=1}^{S} \max |\mathbf{x} - \hat{\mathbf{x}}|
        \label{eq:maxerr}
    \end{equation}
\end{itemize}

\subsection{Performance and accuracy} \label{sec:results}

\subsubsection{Performance Evaluation}

The performance metrics of the VAE and TPM trained on \textit{small}, \textit{medium}, \textit{large} and \textit{extra large} datasets is reported in Tab. \ref{tab:metrics_test} and are evaluated using the MSE, MAE, MRE, and Max Error metrics, as defined in Sec. \ref{section:traintest}. As expected, an increase in training data size enhances both the mean performance metrics and reduces their variance for each models. However, adding more training data benefits the TPM more significantly than the VAE, as the smallest dataset provides only 20 samples for the TPM, which is relatively insufficient to train a robust prediction model. In contrast, for the VAE, these 20 simulations generate 20 $\times N_t$ training samples, offering a substantially larger dataset that is more suitable for its training requirements.

It can be observed that for the \textit{small} and \textit{medium} datasets, the performance of the VAE alone statistically surpasses that of the temporal prediction model. However, as the dataset size increases to the \textit{large} and \textit{extra-large DOE}, the metric distributions of both approaches begin to overlap. While the VAE performance constrains that of the TPM, this is not a theoretical limitation. In principle, the temporal prediction model should be capable of compensating for minor offsets introduced by the VAE as illustrated by the Max Error on the \textit{extra large} database where the temporal prediction surpasses the VAE performance. 

As shown in Tab. \ref{tab:metrics_test}, the maximum error remains important, though it decreases with additional training data. This metric is particularly sensitive to localized high errors; if the model fails to predict the exact position of a shock (characterized by a steep gradient) with pixel-level accuracy, it produces a large local absolute error, resulting in a high maximum error.

As previously discussed in Tab. \ref{tab:datasetssize}, the training time of the ROM model is negligible compared to the time required to produce the database needed to build the model. Therefore, this type of approach is highly beneficial in applications where the model is called significantly more frequently than the number of individuals in the dataset, such as in optimization frameworks or for evaluating unsteady pressure fields under numerous different boundary conditions. These scenarios highlight the substantial potential of such ROMs for efficient and task-specific industrial applications and design process.

\begin{table*}[ht]
\centering
\resizebox{\textwidth}{!}{
\begin{tabular}{| *{14}{c|} }
\hline
\multicolumn{2}{|c|}{ \multirow{2}{*}{\diagbox[width=4.5cm, height=0.82cm]{Models}{Datasets}}} & \multicolumn{3}{c|}{ \textbf{Small} } & \multicolumn{3}{c|}{ \textbf{Medium} } & \multicolumn{3}{c|}{ \textbf{Large} } &   \multicolumn{3}{c|}{ \textbf{Extra Large} }\\
\cline{3-14}
\multicolumn{2}{|c|}{} & \multicolumn{2}{c|}{ $\mu \pm 1.96 \sigma$ } & \textit{Best} & \multicolumn{2}{c|}{ $\mu \pm 1.96 \sigma$ } & \textit{Best} & \multicolumn{2}{c|}{ $\mu \pm 1.96 \sigma$ } & \textit{Best} & \multicolumn{2}{c|}{ $\mu \pm 1.96 \sigma$ } & \textit{Best} \\
 \hline
 \hline
\multirow{4}{*}{\rotatebox{90}{\textbf{VAE}}} & MSE $[kPa^2]$& \multicolumn{2}{c|}{\round{1.696559064388275} $\pm$ \round{0.2756826759358776}} & \round{1.5180315971374512} & \multicolumn{2}{c|}{\round{1.235940682888031} $\pm$ \round{0.170766766830355}} & \round{1.086687684059143} & \multicolumn{2}{c|}{\round{0.9806668296152231} $\pm$ \round{0.1063722718611644}} & \round{0.8961480259895325} & \multicolumn{2}{c|}{\round{0.8415025126934051} $\pm$ \round{0.07348926437055592}} & \round{0.7736866474151611} \\
 & MAE $[kPa]$& \multicolumn{2}{c|}{\round{0.9732597470283508} $\pm$ \round{0.08256419159078646}} &  & \multicolumn{2}{c|}{\round{0.834881911277771} $\pm$  \round{0.057563213959672156}} &  & \multicolumn{2}{c|}{\round{0.7416757661469129} $\pm$ \round{0.0412298819128182}} &  & \multicolumn{2}{c|}{\round{0.6869454658031464} $\pm$ \round{0.02935428503547461}} &  \\
 & MRE $[\%]$ & \multicolumn{2}{c|}{\round{1.5895231887698175} $\pm$ \round{0.1376142669039605}} &  & \multicolumn{2}{c|}{\round{1.3671920988708735} $\pm$ \round{0.09432269463236871}} & & \multicolumn{2}{c|}{\round{1.2164959189843158} $\pm$ \round{0.06693776532102191}} &  & \multicolumn{2}{c|}{\round{1.1243860330432653} $\pm$ \round{0.04761685138857454}} &  \\
  & Mean Max Error $[kPa]$& \multicolumn{2}{c|}{\round{18.743029441833496} $\pm$ \round{1.6336787449577137}} &  & \multicolumn{2}{c|}{\round{16.381305389404297} $\pm$ \round{1.2870744582994935}} & & \multicolumn{2}{c|}{\round{15.04367380726094} $\pm$ \round{1.0782504488555935}} &  & \multicolumn{2}{c|}{\round{14.176535625457763} $\pm$ \round{1.1681156646040238}} &  \\
 \hline
 \hline
 \multirow{5}{*}{\rotatebox{90}{\textbf{TPM}}} & MSE $[kPa^2]$& \multicolumn{2}{c|}{\round{8.451254549026489} $\pm$ \round{2.8221421155085182}} & \round{6.3203253746032715} & \multicolumn{2}{c|}{\round{4.010059833526611} $\pm$ \round{1.5092354974812496}} & \round{2.5878896713256836} & \multicolumn{2}{c|}{\round{1.5568843724879813} $\pm$ \round{0.337455321589564}} & \round{1.1793241500854492} & \multicolumn{2}{c|}{\round{1.0032217979431153} $\pm$ \round{ 0.2891317694339196}} & \round{0.7590394020080566} \\
 & MAE $[kPa]$ & \multicolumn{2}{c|}{\round{2.0531884741783144}  $\pm$ \round{0.34254056918157566}} &  & \multicolumn{2}{c|}{\round{1.4031213593482972} $\pm$ \round{0.26230582905253236}} &  & \multicolumn{2}{c|}{\round{0.8748030256717763} $\pm$ \round{0.08812261201596487}} &  & \multicolumn{2}{c|}{\round{0.7071086704730988} $\pm$ \round{0.0822272994684734}} &  \\
 & MRE $[\%]$ & \multicolumn{2}{c|}{\round{3.3518676534295085} $\pm$ \round{0.5438725673579945}} &  & \multicolumn{2}{c|}{\round{2.31616023555398} $\pm$ \round{0.4296446962389229}} &  & \multicolumn{2}{c|}{\round{ 1.4479712384970898} $\pm$ \round{0.14802476879739063}} & & \multicolumn{2}{c|}{\round{1.1651433072984219} $\pm$ \round{0.1379461043302781}} &  \\
  & Mean Max Error $[kPa]$ &  \multicolumn{2}{c|}{\round{28.533405227661135} $\pm$ \round{3.5766776603954775}}  & & \multicolumn{2}{c|}{\round{23.241562118530272} $\pm$ \round{3.437471000766666}} &  & \multicolumn{2}{c|}{\round{15.342913931988654} $\pm$ \round{1.533422245456512}} &  & \multicolumn{2}{c|}{\round{12.857477493286133} $\pm$ \round{1.4096669688919055}} &  \\
  \cline{3-14}
& Tolerance Outliers $[\%]$ & \multicolumn{3}{c|}{ \roundto{2}{3.2671875}}& \multicolumn{3}{c|}{ \roundto{2}{1.494921875}} & \multicolumn{3}{c|}{ \roundto{2}{0.20781249999999998}} & \multicolumn{3}{c|}{ \roundto{2}{0.080859375}} \\
 \hline

\end{tabular}}
\caption{Performance metric evaluated on the test dataset and proportion of outliers above the tolerance curve discussed in Sec. \ref{sec:FFT}}
\label{tab:metrics_test}
\end{table*}

\subsubsection{Unsteady pressure field predictions}

\begin{figure*}[p]
    \centering
    \begin{subfigure}[b]{0.8\textwidth}
        \centering
        \includegraphics[width=\textwidth, trim = 60 20 60 20]{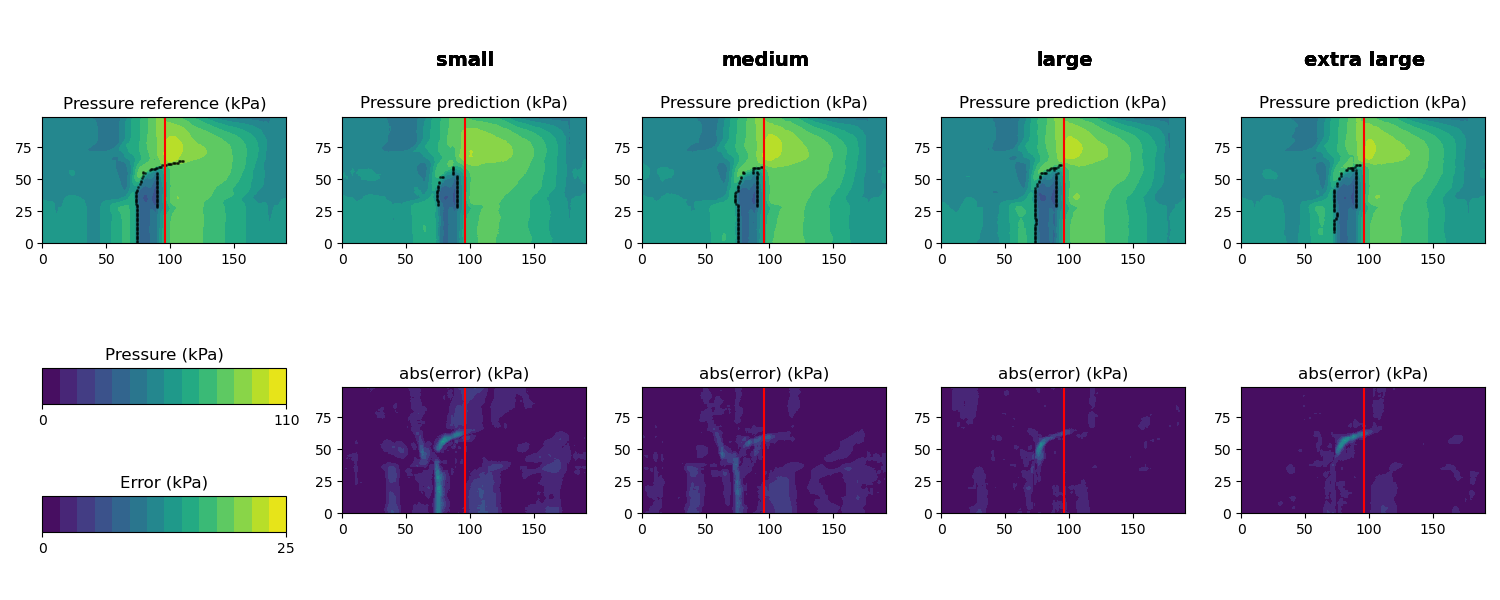}
        \caption{$t = t_s $}
    \end{subfigure}
    \vskip\baselineskip
    \begin{subfigure}[b]{0.8\textwidth}
        \centering
        \includegraphics[width=\textwidth, trim = 60 20 60 20]{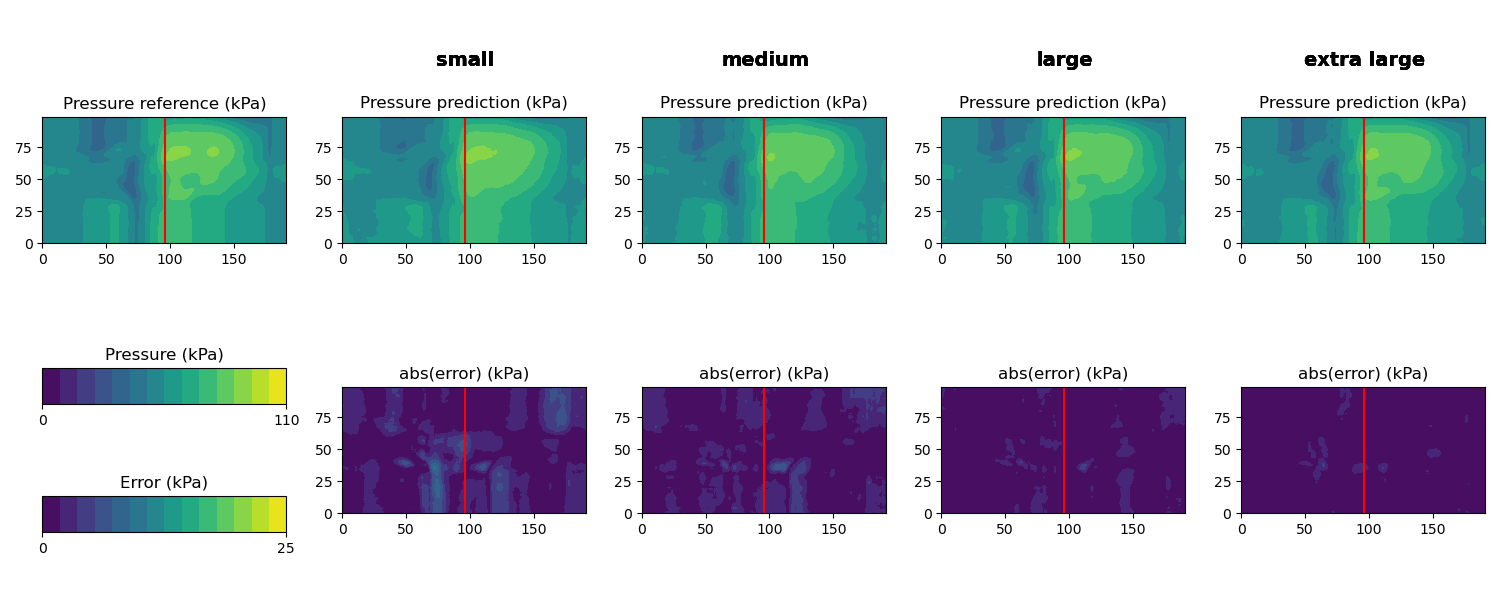}
        \caption{$t = t_s + \frac{1}{3} T_{BPF}$}
    \end{subfigure}
    \vskip\baselineskip
    \begin{subfigure}[b]{0.8\textwidth}
        \centering
        \includegraphics[width=\textwidth, trim = 60 20 60 20]{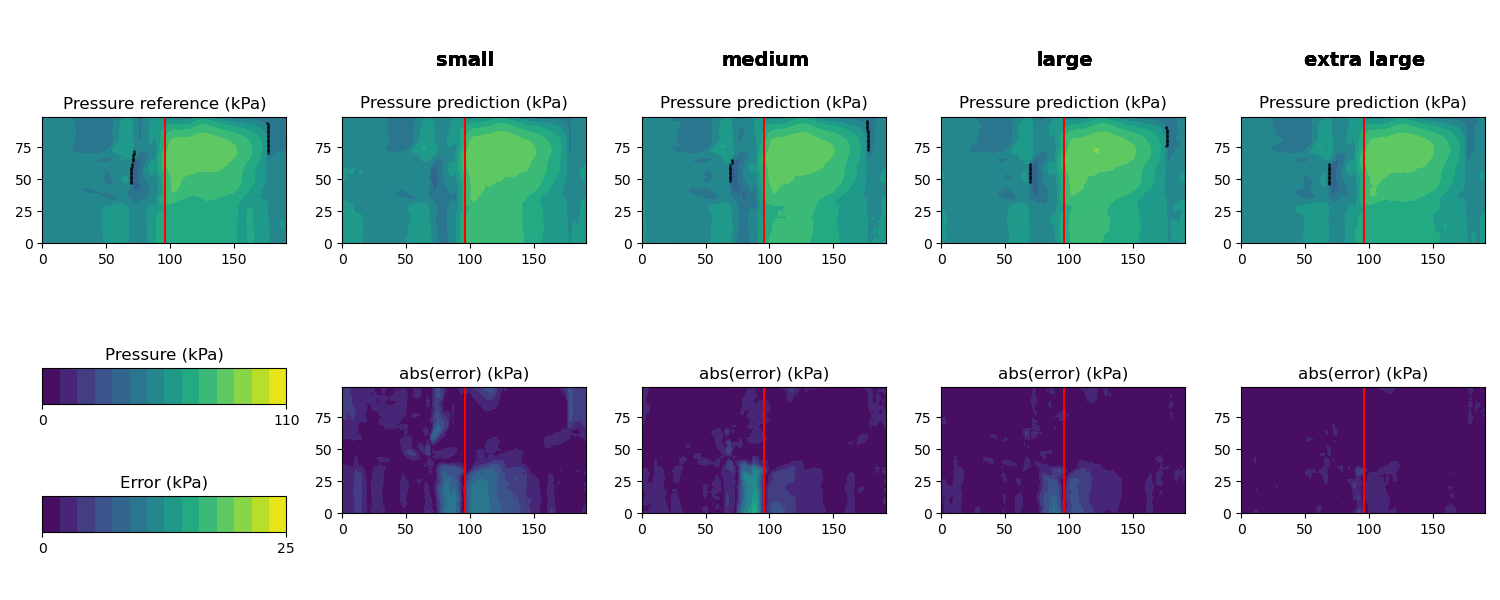}
        \caption{$t = t_s + \frac{2}{3} T_{BPF}$}
    \end{subfigure}

    \caption{Prediction and error of the temporal prediction model for different training datasets and evaluated on \textit{test} dataset at $\alpha = 17.4$, $\beta = 0.6$, $d = 2.61$ and $\Pi = 0.29$}
    \label{fig:fullpage}
\end{figure*}

\begin{figure*}[p]
    \centering
    \begin{subfigure}[b]{0.8\textwidth}
        \centering
        \includegraphics[width=\textwidth, trim = 60 20 60 20]{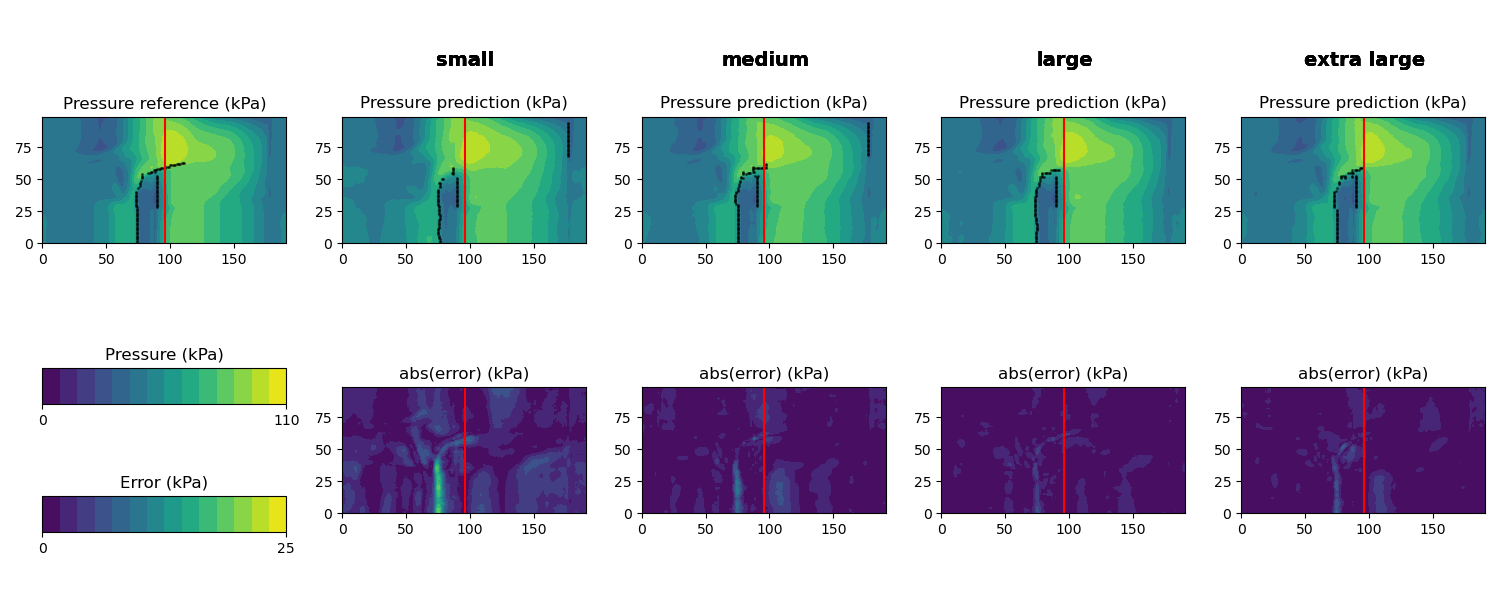}
        \caption{$t = t_s $}
    \end{subfigure}
    \vskip\baselineskip
    \begin{subfigure}[b]{0.8\textwidth}
        \centering
        \includegraphics[width=\textwidth, trim = 60 20 60 20]{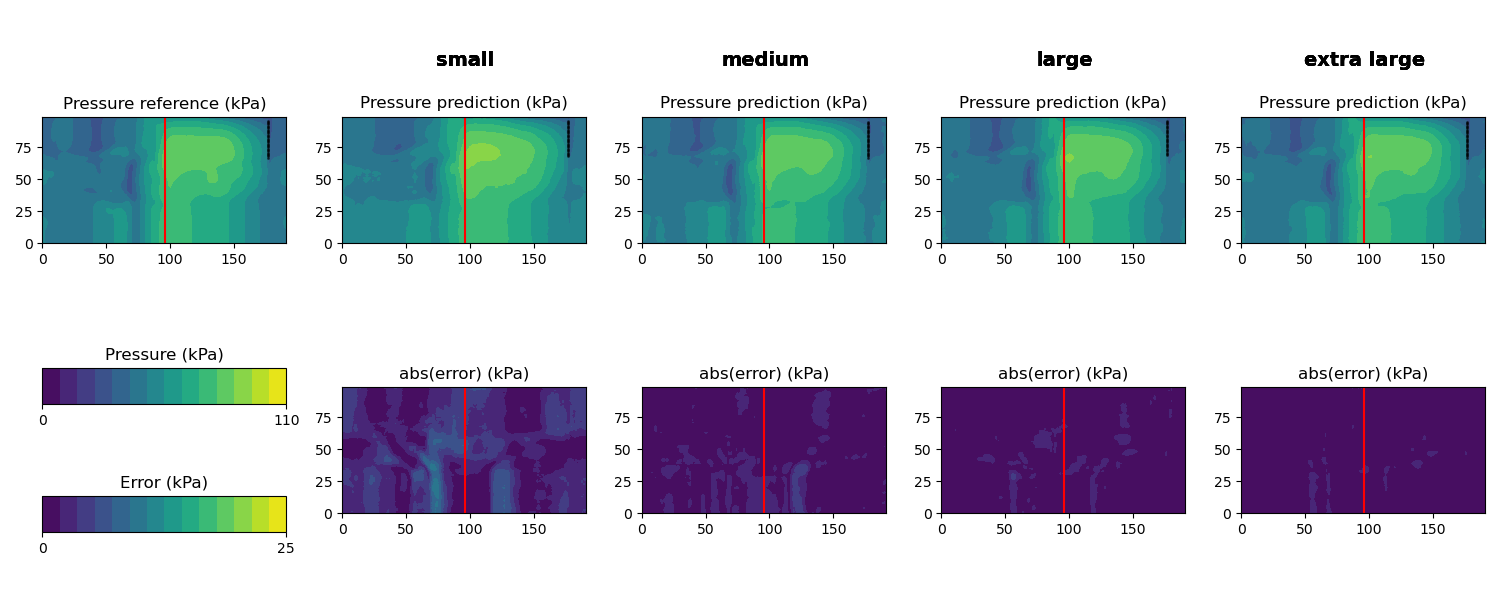}
        \caption{$t = t_s + \frac{1}{3} T_{BPF}$}
    \end{subfigure}
    \vskip\baselineskip
    \begin{subfigure}[b]{0.8\textwidth}
        \centering
        \includegraphics[width=\textwidth, trim = 60 20 60 20]{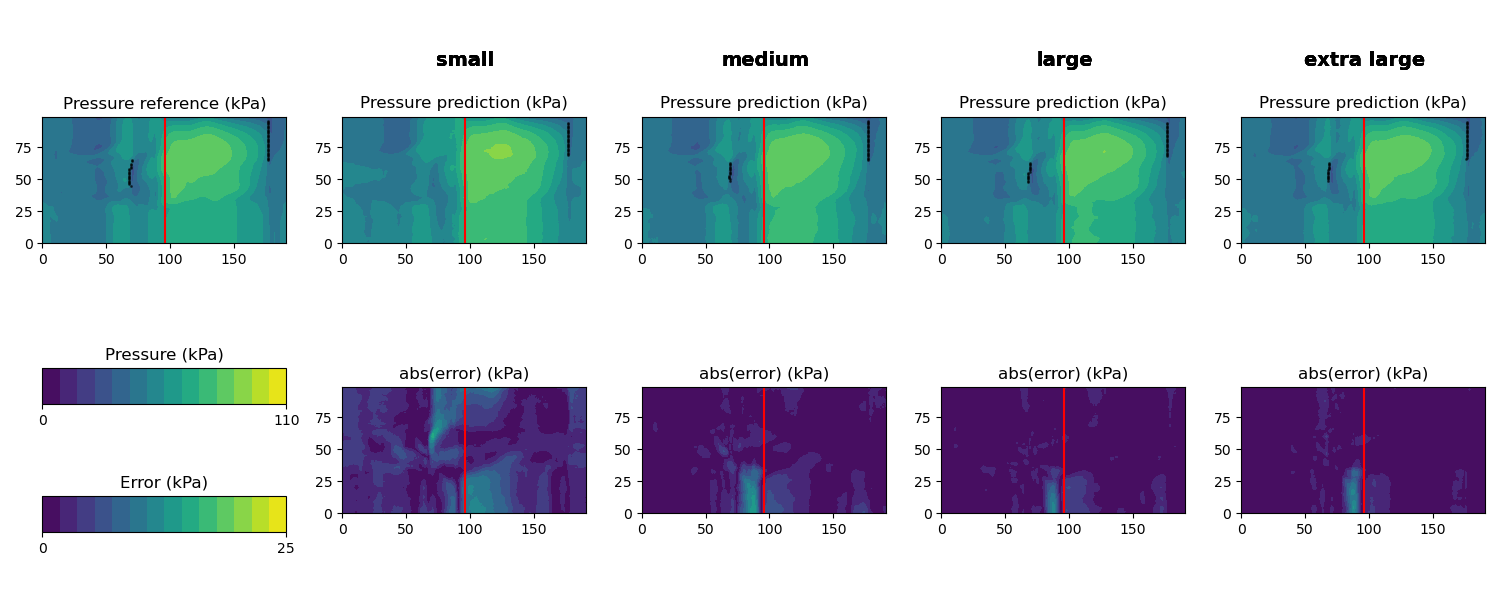}
        \caption{$t = t_s + \frac{2}{3} T_{BPF}$}
    \end{subfigure}

    \caption{Prediction and error of the temporal prediction model for different training datasets and evaluated on \textit{test} dataset at $\alpha = -18.6$, $\beta = 0.05$, $d = 1.19$ and $\Pi = 0.28$}
    \label{fig:fullpage2}
\end{figure*}

Figures \ref{fig:fullpage} and  \ref{fig:fullpage2} illustrate the pressure field predictions produced by the temporal model at different time steps across various simulations within the test set. Even when trained on the \textit{small} dataset, the pressure predictions are already reasonably accurate, capturing the overall structure of the pressure fields within less than 5\% uncertainty. However, reconstruction errors remain notable for finer features and at pressure field discontinuities. This result illustrates the model robustness and its ability to achieve satisfactory reconstruction quality even with limited data (20 simulations for the smallest dataset). As dataset size increases, the model more accurately captures detailed features, including shock patterns, resulting in significantly reduced reconstruction errors when trained on the largest dataset. 

Temporal stability is often a challenge for machine learning models predicting time sequences. However, as shown in Fig. \ref{fig:fullpage} and  \ref{fig:fullpage2}, the TPM error remains stable over time, without drifting, even in low-data scenarios.

Finally, the largest errors consistently occur near shocks, indicated as Canny edge-detected discontinuities in the reference pressure fields. As expected, the inclusion of additional data progressively reduces error magnitude in these regions, leading to increasingly accurate shock reconstruction.

 \begin{figure}[htbp]
    \centering
    \includegraphics[width=0.60\textwidth]{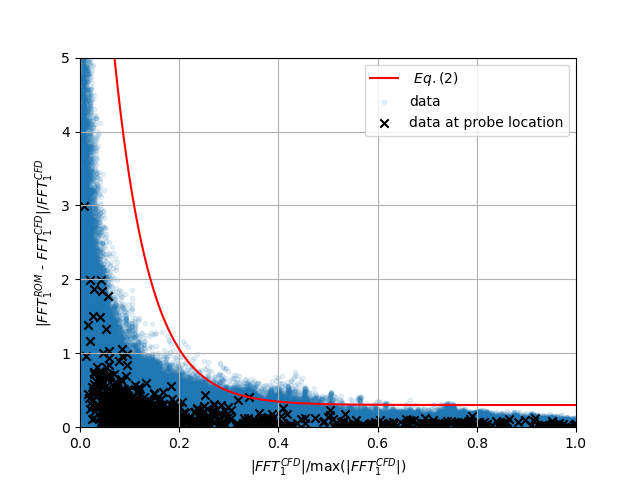}
    \caption{Evaluation of the accuracy of the first harmonic of the Fourier transform of the predicted pressure field of the best model on the \textit{large} dataset compared to the reference simulated field. The red line is the experimentally-based tolerance curve and the black crosses are indicating points where experimental data is available.} 
    \label{fig:tolerance}
\end{figure} 

\subsubsection{Amplitude of Fourier harmonics} \label{sec:FFT}

As described in Sec. \ref{section:GAF}, industrial mechanical design often prioritizes the first harmonic of the Fourier transform pressure fluctuations. The time signal used for Fourier analysis consists of $N_t$ samples extracted over the time period $T_{BPF}$. Consequently, the first harmonic of the Fourier transform corresponds to the fundamental frequency of a rotor blade passing through a stator channel. Here, model predictions were compared against an experimentally derived tolerance curve from Eq. \ref{eq:tolerance}, which represents an acceptable range for such quantity. The parameters of this so-called tolerance curve were calibrated based on experimental measurements provided by a limited number of probes and at specific wind tunnel operating conditions.

To ensure a fair comparison, model performance was initially evaluated exclusively at the experimental probe locations. In Fig. \ref{fig:tolerance}, the tolerance curve derived from experimental data is displayed in red, with black crosses representing the TPM first harmonic amplitude at these probe points. For the \textit{large} dataset, all black crosses are located below the tolerance curve, indicating that the model meets the experimental quality criteria.

Blue points in Fig. \ref{fig:tolerance} denote the temporal model first harmonic amplitude across the entire blade surface. While most of these points fall below the tolerance curve, a small portion of outliers remains. Table \ref{tab:metrics_test} further quantifies this result by reporting the proportion of points exceeding the tolerance curves, named Tolerance Outliers, for different dataset sizes. With larger datasets, this proportion decreases rapidly, indicating that adherence to the tolerance curve improves significantly with additional training data, particularly for the largest datasets.

\section{Conclusion}

This study presents a deep-learning-based Reduced Order Model (ROM) for predicting unsteady pressure fields on turbine rotor blades. By combining a Variational Auto-Encoder decoder with a Gated Recurrent Unit, the model effectively captures complex nonlinear phenomena, including moving shocks, and provides robust predictions across various boundary conditions.

Key findings indicate that the ROM achieves satisfactory accuracy even with limited training data, making it a valuable tool for early-stage design evaluations. Its ability to approximate the first harmonic components of the Fourier transform suggests potential for predicting Generalized Aerodynamic Forces. However, the associated projection of pressure predictions onto mechanical deformation modes adds complexity to the model, requiring further work to ensure reliable predictions.

Additionally, we quantified the influence of training dataset size on model performance, recognizing that database size is a crucial factor for practical industrial deployment. Specifically, a dataset of 20 individuals is sufficient for reconstructing the pressure field with uncertainties below 5\%. However, accurately capturing the first harmonic of the Fourier transform requires a larger dataset. Our results show that the number of tolerance outliers is already acceptable with 40 individuals. To achieve uncertainty levels comparable to those observed between simulations and experimental data, a dataset of 80 individuals is required. Although overall accuracy improves with dataset size, the maximum error is less responsive to increases in number of individuals. This highlights the challenge of predicting shock locations, where small spatial errors can lead to significant local discrepancies. These findings provide a preliminary assessment of the achievable accuracy relative to dataset size, which can be further optimized in future work to meet specific application requirements. Such optimization may involve advanced training strategies or tailored model architectures to improve shock localization and overall predictive precision.

This work serves as a proof of concept, demonstrating the potential of nonlinear ROM methodologies to reduce the computational cost associated with running numerous URANS simulations. However, the cost of building the simulation database remains significant. Given that model training time is negligible in comparison, this approach is most effective when the model is used more frequently than the number of URANS simulations required to build the database. This balance also depends on the desired accuracy of the model. Potential applications include optimization frameworks and evaluations across multiple boundary conditions, where the rapid deployment of the ROM can significantly accelerate decision-making processes.

\section*{Acknowledgments}

The presented work has been performed in the framework of the HE-ART project, part of the Clean Aviation Joint Undertaking, and funded by the European Union’s Horizon 2020 research and innovation programme under grant agreement N◦ 101102013. It has also benefited from computational resources made available on the Tier-1 supercomputer of the Fédération Wallonie-Bruxelles, infrastructure funded by the Walloon Region under grant agreement N◦ 1910847.

\bibliographystyle{unsrt}  
\bibliography{references}

\begin{thebibliography}{10}

\bibitem{winter2016}
Maximilian Winter and Christian Breitsamter.
\newblock Efficient unsteady aerodynamic loads prediction based on nonlinear system identification and proper orthogonal decomposition.
\newblock {\em Journal of Fluids and Structures}, 67:1--21, 2016.

\bibitem{karbasian2022}
Hamid~R Karbasian and Brian~C Vermeire.
\newblock Application of physics-constrained data-driven reduced-order models to shape optimization.
\newblock {\em Journal of Fluid Mechanics}, 934:A32, 2022.

\bibitem{hasegawa2020}
Kazuto Hasegawa, Kai Fukami, Takaaki Murata, and Koji Fukagata.
\newblock Machine-learning-based reduced-order modeling for unsteady flows around bluff bodies of various shapes.
\newblock {\em Theoretical and Computational Fluid Dynamics}, 34:367--383, 2020.

\bibitem{rozov2021}
Vladyslav Rozov and Christian Breitsamter.
\newblock Data-driven prediction of unsteady pressure distributions based on deep learning.
\newblock {\em Journal of Fluids and Structures}, 104:103316, 2021.

\bibitem{ma2022}
Jiaobin Ma, Zhufeng Liu, Yunzhu Li, and Yonghui Xie.
\newblock Prediction method of unsteady flow load of compressor stator under working condition disturbance.
\newblock {\em Applied Sciences}, 12(22):11566, 2022.

\bibitem{hines2023}
Derrick Hines and Philipp Bekemeyer.
\newblock Graph neural networks for the prediction of aircraft surface pressure distributions.
\newblock {\em Aerospace Science and Technology}, 137:108268, 2023.

\bibitem{solera2024}
Alberto Solera-Rico, Carlos Sanmiguel~Vila, Miguel G{\'o}mez-L{\'o}pez, Yuning Wang, Abdulrahman Almashjary, Scott~TM Dawson, and Ricardo Vinuesa.
\newblock $\beta$-variational autoencoders and transformers for reduced-order modelling of fluid flows.
\newblock {\em Nature Communications}, 15(1):1361, 2024.

\bibitem{zhang2022}
Chao Zhang and Matthew Janeway.
\newblock Optimization of turbine blade aerodynamic designs using cfd and neural network models.
\newblock {\em International Journal of Turbomachinery, Propulsion and Power}, 7(3):20, 2022.

\bibitem{nigro2025journal}
R{\'e}my Nigro, Lieven Baert, Florence Nyssen, Jean de~Cazenove, Joachim Dominique, Ingrid Lepot, Monica Veglio, and R{\'e}my Princivalle.
\newblock Multifidelity aeromechanical design framework for high flow speed multistage axial compressors.
\newblock {\em Journal of Turbomachinery}, 147(4), 2025.

\bibitem{kingma2013}
Diederik~P Kingma.
\newblock Auto-encoding variational bayes.
\newblock {\em arXiv preprint arXiv:1312.6114}, 2013.

\bibitem{denos2005}
R~D{\'e}nos and G~Paniagua.
\newblock Effect of vane-rotor interaction on the unsteady flowfield downstream of a transonic high pressure turbine.
\newblock {\em Proceedings of the Institution of Mechanical Engineers, Part A: Journal of Power and Energy}, 219(6):431--442, 2005.

\bibitem{paniagua2008}
Guillermo Paniagua, Tolga Yasa, Adres de~la Loma, Lionel Castillon, and Thomas Coton.
\newblock Unsteady strong shock interactions in a transonic turbine: experimental and numerical analysis.
\newblock {\em Journal of propulsion and power}, 24(4):722--731, 2008.

\bibitem{laumert2002_part1}
Bjorn Laumert, Hans Martensson, and Torsten~H Fransson.
\newblock Investigation of unsteady aerodynamic blade excitation mechanisms in a transonic turbine stage—part i: Phenomenological identification and classification.
\newblock {\em J. Turbomach.}, 124(3):410--418, 2002.

\bibitem{laumert2002_part2}
Bjorn Laumert, Hans Martensson, and Torsten~H Fransson.
\newblock Investigation of unsteady aerodynamic blade excitation mechanisms in a transonic turbine stage—part ii: Analytical description and quantification.
\newblock {\em J. Turbomach.}, 124(3):419--428, 2002.

\bibitem{gerolymos2002}
GA~Gerolymos, GJ~Michon, and Julien Neubauer.
\newblock Analysis and application of chorochronic periodicity in turbomachinery rotor/stator interaction computations.
\newblock {\em Journal of propulsion and power}, 18(6):1139--1152, 2002.

\bibitem{cambier2011}
L~Cambier, M~Gazaix, S~Heib, S~Plot, M~Poinot, JP~Veuillot, JF~Boussuge, and M~Montagnac.
\newblock An overview of the multi-purpose elsa flow solver.
\newblock {\em Aerospace Lab}, (2):p--1, 2011.

\bibitem{giles1990}
Michael~B Giles.
\newblock Stator/rotor interaction in a transonic turbine.
\newblock {\em Journal of Propulsion and Power}, 6(5):621--627, 1990.

\bibitem{canny1986}
John Canny.
\newblock A computational approach to edge detection.
\newblock {\em IEEE Transactions on pattern analysis and machine intelligence}, (6):679--698, 1986.

\bibitem{fujimoto2019}
Takeshi~R Fujimoto, Taro Kawasaki, and Keiichi Kitamura.
\newblock Canny-edge-detection/rankine-hugoniot-conditions unified shock sensor for inviscid and viscous flows.
\newblock {\em Journal of Computational Physics}, 396:264--279, 2019.

\bibitem{znamenskaya2021}
Irina~A Znamenskaya and Igor~A Doroshchenko.
\newblock Edge detection and machine learning for automatic flow structures detection and tracking on schlieren and shadowgraph images.
\newblock {\em Journal of Flow Visualization and Image Processing}, 28(4), 2021.

\bibitem{doroshchenko2023}
Igor~A Doroshchenko.
\newblock Analysis of the experimental flow shadowgraph images by computer vision methods.
\newblock {\em Numerical Methods and Programming (Vychislitel'nye Metody i Programmirovanie)}, 24(2):231--242, 2023.

\bibitem{dowell2021modern}
Earl~H Dowell.
\newblock {\em A modern course in aeroelasticity}, volume 264.
\newblock Springer Nature, 2021.

\bibitem{SafranHE}
Jacques Demolis.
\newblock Safran helicopter engines - document interne.
\newblock 2024.

\bibitem{romero2006}
Vicente~J Romero, John~V Burkardt, Max~D Gunzburger, and Janet~S Peterson.
\newblock Comparison of pure and “latinized” centroidal voronoi tessellation against various other statistical sampling methods.
\newblock {\em Reliability Engineering \& System Safety}, 91(10-11):1266--1280, 2006.

\bibitem{sainvitu2010}
Caroline Sainvitu, Vicky Iliopoulou, and Ingrid Lepot.
\newblock Global optimization with expensive functions-sample turbomachinery design application.
\newblock In {\em Recent Advances in Optimization and its Applications in Engineering: The 14th Belgian-French-German Conference on Optimization}, pages 499--509. Springer, 2010.

\bibitem{beaucaire2019}
Paul Beaucaire, Ch~Beauthier, and Caroline Sainvitu.
\newblock Multi-point infill sampling strategies exploiting multiple surrogate models.
\newblock In {\em Proceedings of the Genetic and Evolutionary Computation Conference Companion}, pages 1559--1567, 2019.

\bibitem{zhang2005}
Tong Zhang and Bin Yu.
\newblock Boosting with early stopping: Convergence and consistency.
\newblock {\em The Annals of Statistics}, 33(4), August 2005.

\bibitem{van2008visualizing}
Laurens Van~der Maaten and Geoffrey Hinton.
\newblock Visualizing data using t-sne.
\newblock {\em Journal of machine learning research}, 9(11), 2008.

\bibitem{pytorch}
Adam Paszke, Sam Gross, Francisco Massa, Adam Lerer, James Bradbury, Gregory Chanan, Trevor Killeen, Zeming Lin, Natalia Gimelshein, Luca Antiga, et~al.
\newblock Pytorch: An imperative style, high-performance deep learning library.
\newblock {\em Advances in neural information processing systems}, 32, 2019.

\bibitem{higgins2017}
Irina Higgins, Loic Matthey, Arka Pal, Christopher~P Burgess, Xavier Glorot, Matthew~M Botvinick, Shakir Mohamed, and Alexander Lerchner.
\newblock beta-vae: Learning basic visual concepts with a constrained variational framework.
\newblock {\em ICLR (Poster)}, 3, 2017.

\bibitem{burgess2018understanding}
Christopher~P Burgess, Irina Higgins, Arka Pal, Loic Matthey, Nick Watters, Guillaume Desjardins, and Alexander Lerchner.
\newblock Understanding disentangling in $\beta$-vae.
\newblock {\em arXiv preprint arXiv:1804.03599}, 2018.

\bibitem{chung2014empirical}
Junyoung Chung, Caglar Gulcehre, KyungHyun Cho, and Yoshua Bengio.
\newblock Empirical evaluation of gated recurrent neural networks on sequence modeling.
\newblock {\em arXiv preprint arXiv:1412.3555}, 2014.

\bibitem{dey2017gate}
Rahul Dey and Fathi~M Salem.
\newblock Gate-variants of gated recurrent unit (gru) neural networks.
\newblock In {\em 2017 IEEE 60th international midwest symposium on circuits and systems (MWSCAS)}, pages 1597--1600. IEEE, 2017.

\bibitem{zimmermann2012forecasting}
Hans-Georg Zimmermann, Christoph Tietz, and Ralph Grothmann.
\newblock Forecasting with recurrent neural networks: 12 tricks.
\newblock {\em Neural Networks: Tricks of the Trade: Second Edition}, pages 687--707, 2012.

\end{thebibliography}

\end{document}